\documentstyle{article}[12pt]\begin{document}\centerline{\bf Some Integrals of the Dedekind $\eta$-Function}\vskip .5in
\centerline{M.L. Glasser}
\centerline{Department of Physics}
\centerline{Clarkson University, Potsdam, NY 13699-5280(USA)}\vskip .5in
\centerline{e-mail: laryg@clarkson.edu}
\centerline{Abstract}
\begin{quote}
\noindent
This note presents selected values of the class of integrals
$$\int_0^{\infty}f(x)\eta^n(ix)dx.$$
\end{quote}
\vskip 1in
\noindent
Keywords: Ramanujan, Glaisher, integral, Dedekind-eta q-series

\vskip 1in
\noindent
2000 Mathematics Classification: Primary 33E05, 44A20; secondary 44A10

\newpage

The motivation for this study was to investigate integrals similar to one that Ramanujan included as item IV(5) in his famous 1913 letter to Hardy [1] containing the expression $\cosh(x)+\cos(x)$ in the denominator. Actually, the first to examine an integral of this class appears to have been J.W.L. Glaisher[2] in 1871; it seems doubtful, however, that Ramanujan would have been aware of this work. The study of these integrals quickly led to  the problem of evaluating integrals of the Dedekind Eta function, and particularly  pertaining to the class
\begin{eqnarray} \int_0^{\infty}f(x)\eta^n(ix)dx\label{eqn1}\end{eqnarray}
This is probably to have been expected, since Ramanujan devotes  a good deal of space in his notebooks to such integrals. These and many others can be found in Berndt's seminal exposition of Ramanujan's work[3]. Other Dedekind function integrals appear scattered in the literature, such as in references [4,5] and works cited there. The aim of this paper is to present a number of results of the form (1) having a more elementary character.  

 Note  that  for the real nome $q=e^{-2\pi x}$, our instance of the Dedekind $\eta$-function is given by
\begin{eqnarray}\eta(ix)=q^{1/24}\prod_{n=1}^{\infty}(1-q^n).\label{eqn2}\end{eqnarray}
We begin with a simple rearrangement of Euler's  identity[6]
\begin{eqnarray}\eta(ix)=\frac{2}{\sqrt{3}}\sum_{n=0}^{\infty}\cos[(2n+1)\pi/6]q^{(2n+1)^2/24}.\label{eqn3}\end{eqnarray}
Then, after a change of variable $q=e^{-2\pi x}$, one has
\begin{eqnarray}\int_0^1\frac{dq}{q}q^y\eta(ix)=\frac{2\pi}{\sqrt{3}}\sum_{n=0}^{\infty}\cos[(2n+1)\frac{\pi}{6}]\int_0^{\infty}dx\; e^{-2\pi[y+(2n+1)^2/24]x}\nonumber\\
=\frac{48\pi}{\sqrt{3}}\sum_{n=0}^{\infty}\frac{\cos[(2n+1)\pi/6]}{(2n+1)^2+24y}=\pi\sqrt{\frac{2}{y}}\frac{\sinh\pi\sqrt{8y/3}}{\cosh\pi\sqrt{6y}}.\label{eqn4}\end{eqnarray}
We therefore have the Laplace transform, where $ t=3\pi y$,
\begin{eqnarray}\int_0^{\infty}e^{-xt}\eta(ix)dx=\sqrt{\frac{\pi}{t}}\frac{\sinh2\sqrt{\pi t/3}}{\cosh\sqrt{3\pi t}}.\label{eqn5}\end{eqnarray}
In what follows, we shall assume that all parameters take on real values for which the integrals converge absolutely.
\vskip .1in

Let $F(t)$ denote the inverse Laplace transform of $f(x)$. Then from (5)  by multiplying both sides by $F(t)$ and integrating over $t$,we have, by invoking Fubini's theorem, the identity
\begin{eqnarray}\int_0^{\infty}f(x)\eta(ix)dx=\sqrt{\pi}\int_0^{\infty}\frac{dt}{\sqrt{t}}F(t)\frac{\sinh2\sqrt{\pi t/3}}{\cosh\sqrt{3\pi t}}. \label{eqn6}\end{eqnarray}

We obtain in this way[8]
\begin{eqnarray}\int_0^{\infty}x^{-s}\eta(ix)dx=\mbox{\hskip 2in}\nonumber\\
\frac{8\sqrt{3}\pi}{16^s(3\pi)^s}\frac{\Gamma(2s-1)}{\Gamma(s)}[\zeta(2s-1,\frac{1}{12})+\zeta(2s-1,\frac{11}{12})\nonumber\\-\zeta(2s-1,\frac{5}{12})-\zeta(2s-1,\frac{7}{12})],\mbox{\hskip .2in}(s>0).\end{eqnarray}
Next, in (5) replace $t$ by $it$ and formally take the real part of both sides to get
\begin{eqnarray}\int_0^{\infty}\cos(xy)\eta(ix)dx=\sqrt{\frac{\pi}{2y}}\frac{\sinh\sqrt{8\pi y/3}+\sin\sqrt{8\pi y/3}}{\cosh\sqrt{8\pi y/3}+\cos\sqrt{8\pi y/3}}.\end{eqnarray}
For $y\rightarrow0$, (8) becomes 
\begin{eqnarray}\int_0^{\infty}\eta(ix)dx=\frac{2\pi}{\sqrt{3}}.\end{eqnarray}
Similarly,
\begin{eqnarray}\int_0^{\infty}\sin(xy)\eta(ix)dx=\sqrt{\frac{\pi}{2y}}\frac{\sinh\sqrt{8\pi y/3}-\sin\sqrt{8\pi y/3}}{\cosh\sqrt{8\pi y/3}+\cos\sqrt{8\pi y/3}}.\end{eqnarray}
By dividing both sides of (10) by $y$ and integrating over y from 0 to $\infty$, we obtain
\begin{eqnarray}\int_0^{\infty}\frac{dx}{x^2}\frac{\sinh(x)-\sin(x)}{\cosh(x)+\cos(x)}=\frac{\pi}{4}\end{eqnarray}
an integral of the type that led to this study.

\vskip .1in
Next, Jacobi's triple identity [7] may be written
\begin{eqnarray} \eta^3(ix)=\sum_{n=0}^{\infty}(-1)^n(2n+1)q^{(2n+1)^2/8}.\end{eqnarray}
By multiplying both sides by $q^{z-1}$ and integrating over $0<q<1$ as before, we find, with $q=e^{-2\pi x}$, 
\begin{eqnarray}
\int_0^1\frac{dq}{q}q^z\eta^3(ix)=8\sum_{n=0}^{\infty}\frac{(-1)^n(2n+1)}{(2n+1)^2+8z}=\frac{2\pi}{\cosh(\pi\sqrt{2z})},\end{eqnarray}

which gives us the Laplace transform
\begin{eqnarray}
\int_0^{\infty}e^{-xy}\eta^3(ix)dx=\rm{sech}\sqrt{\pi y}.\end{eqnarray}
As for (6), if $F(t)$ denotes the inverse Laplace transform of $f(x)$, we  have the identity
\begin{eqnarray}\int_0^{\infty}f(x)\eta^3(ix)dx=\int_0^{\infty}\frac{F(t)}{\rm{cosh}\sqrt{\pi t}}dt.\end{eqnarray}

Formula (15) is much more flexible than  (6) and leads to a variety of interesting looking integrals, a number of which are displayed in the appendix, including the mysterious (A7)
\begin{eqnarray}\int_0^{\infty}\sqrt{\frac{\sqrt{x^2+1}-1}{x^2+1}}e^{-\pi x/4}\prod_{n=0}^{\infty}(1-e^{-2\pi nx})^3dx=\sqrt{2}-1.\end{eqnarray}
This is derived by noting that the Laplace transform of $F(t)=\sin(at)/\sqrt{\pi t}$ is the Imaginary part of $(a+i x)^{-1/2}$. Inserting this into (15) and using Parseval's identity for the cosine Fourier transform to simplify the integral on the right hand side, one obtains (16) for $a=1$

By proceeding analogously to the derivation of (11), we find a second example of the ``Glaisher-Ramanujan" class
\begin{eqnarray}\int_0^{\infty}\frac{dx}{x}\frac{\sinh(x/2)\sin(x/2)}{\cosh(x)+\cos(x)}=\frac{\pi}{8}.\end{eqnarray}
\vskip .1in

In conclusion, we have opened a way to produce many integrals over the Dedekind-$\eta$ function and evaluated one or two Glaisher-Ramnujan integrals. But, there exist many further $q-$ identities similar to (3) and (12) which might prove productive for extending this investigation.

\vskip .3in
\noindent{\bf Acknowledgment} 

The author thanks the Department of Theoretical, Atomic and Optical Physics, Universidad de Valladolid, for generous hospitality while this work was completed.

\newpage

\centerline{\bf Appendix}\vskip .2in

$$\int_0^{\infty}e^{-xy}\eta^3(ix){\,dx}={\rm sech}\sqrt{\pi y}\eqno(A1)$$
$$\int_0^{\infty}\frac{\eta^3(ix)}{x+a}{\,dx}=\frac{2}{\pi}\int_0^{\infty}\frac{xe^{-ax^2/\pi}}{\cosh(x)}{\,dx}\eqno(A2)$$
$$\int_0^{\infty}x^{-\nu}\eta^3(ix){\,dx}=\frac{4}{\pi^{\nu}}\frac{\Gamma(2\nu)}{\Gamma(\nu)}\beta(2\nu)\mbox{\hskip .2in}(\nu>0)\eqno(A3)$$
$$\int_0^{\infty}\frac{\eta^3(ix)}{\sqrt{x+a}}{\,dx}=\frac{2}{\pi}\int_0^{\infty}\frac{e^{-ax^2/\pi}}{\cosh(x)}{\,dx}\eqno(A4)$$
$$\int_0^{\infty}x^{-1/2}e^{-a/x}\eta^3(ix){\,dx}={\rm sech}\sqrt{\pi a}\eqno(A5)$$

$$\int_0^{\infty}e^{-xy}\eta^3(ix)\frac{{\,dx}}{x}=\frac{2}{\pi}\int_{\sqrt{\pi y}}^{\infty}x\,{\rm sech}(x){\,dx}\eqno(A6)$$

$$\int_0^{\infty}\sqrt{\frac{\sqrt{x^2+1}-1}{x^2+1}}\,\eta^3(ix){\,dx}=\sqrt{2}-1\eqno(A7)$$
$$\int_0^{\infty}x^{-1/2}\cos(a/x)\eta^3(ix){\,dx}=2\,\frac{\cos\sqrt{\pi a/2}
\cosh\sqrt{\pi a/2}}{\cos\sqrt{2\pi a}+\cosh\sqrt{2\pi a}}\eqno(A8)$$
$$\int_0^{\infty}x^{-1/2}{\rm erf}(\sqrt{bx})\eta^3(ix){\,dx}=\frac{4}{\pi}\arctan(\tanh\frac{1}{2}\sqrt{\pi b})\eqno(A9)$$
$$\int_0^{\infty}x^{-1/2}e^{a/x}{\rm erfc}(\sqrt{a/x})\eta^3(ix){\,dx}=$$
$$\frac{1}{\pi\sqrt{a}}\left[\psi(\frac{1}{2}\sqrt{a/\pi}+\frac{3}{4})-\psi(\frac{1}{2}\sqrt{a/\pi}+\frac{1}{4})\right]\eqno(A10)$$
$$\int_0^{\infty}\cos(xy)\eta^3(ix)dx=\frac{\cosh(\sqrt{\pi y/2})\cos(\sqrt{\pi y/2})}{\sinh^2(\sqrt{\pi y/2})+\cos^2(\sqrt{\pi y/2})}\eqno(A11)$$
$$\int_0^{\infty}\sin(xy)\eta^3(ix)dx=\frac{\sinh(\sqrt{\pi y/2})\sin(\sqrt{\pi y/2})}{\sinh^2(\sqrt{\pi y/2})+\cos^2(\sqrt{\pi y/2})}\eqno(A12)$$
$$\int_0^{\infty}\eta(ix)dx=\frac{2\pi}{\sqrt{3}}\eqno(A13)$$
$$\int_0^{\infty}\eta^3(ix)dx=1\eqno(A14)$$
$$\int_0^{\infty}x^n\eta^3(ix)dx=\frac{4n!}{\pi^{n+1}}\beta(2n+1)\eqno(A15)$$

\newpage
\centerline{References}\vskip .1in
\noindent
[1] G.H. Hardy, {\it Ramanujan}, Cambridge University Press, 1940.

\noindent
[2] J.W.L. Glaisher,{\it  On the Summation by Definite Integrals of Geometric Series of the Second and Higher Order},  Quarterly J. Math. (Oxford Series 11) 328 (1871).  Equation (34) of this paper is unfortunately incorrect.

\noindent
[3] B.C. Berndt, {\it Ramanujan's Notebooks, Part III}, Springer-Verlag, New York, 1991.

\noindent
[4]  Liang-Cheng Zhang, {\it  Some q-integrals Associated with Modular Forms},  J. Math. Anal. and Appl.
{\bf{150}}, 264-273 (1990).

\noindent
[5]  S.H. Son, {\it Some Integrals of Theta Functions in Ramanujan's Lost Notebook},  in {\it Fifth Conference of the Canadian Number Theory Association}. (R. Gupta and K.S. Williams, eds.) CRM  Proc. and Lecture Notes, Vol. 19 American Mathematical Society, Providence, RI. 1999, pp.323-332.

\noindent
[6] Nathan J. Fine,{\it Basic Hypergeometric Series and Applications}, Mathematical Surveys and Monographs, Number 27, American Mathematical Society, Providence, R.I., 1988. p.7.

\noindent
[7] C.G.I.  Jacobi, {\it Fundamenta Nova Theoriae Functionam Ellipticarum }, Konigsberg (1829), p.186.

\noindent
[8] A.P. Prudikov et al. {Integrals and Series, Vol. i}, Nauka Publishers, Moscow, 1981. Eq.(2.4.5(4)). This reference contains all the other definite integrals used in this study.
\end{document}